\documentclass[12pt]{article}

\usepackage[dvips]{graphicx}
\usepackage[dvips]{color}

\usepackage{amssymb,wrapfig}
\usepackage{amsmath}
\usepackage{array,longtable}

\newcolumntype{V}{>{$}m{4cm}<{$}}
\newcolumntype{C}{>{$}c<{$}}
\newcolumntype{L}{>{$}l<{$}}
\newcolumntype{R}{>{$}r<{$}}

\newcommand{\pd}{\partial}

\usepackage{hyperref}
\renewcommand{\theequation}{\arabic{section}.\arabic{equation}}

\begin{document}
\title{
\textsc{Tachyon Condensation}\\
\textsc{in}\\
\textsc{Cubic Superstring Field Theory}\\}
\author{
\textsf{I.Ya. Aref'eva\footnote{Email: arefeva@mi.ras.ru},
\,A.S. Koshelev\footnote{Email: kas@depni.npi.msu.su}}\\
\emph{Steklov Mathematical Institute,}\\
\emph{Gubkin st. 8, Moscow, Russia, 117966}\\
\\
\textsf{D.M. Belov\footnote{Email: belov@orc.ru}}\\
\emph{Physical Department, Moscow State University,}\\
\emph{Moscow, Russia, 119899}\\
\\
\textsf{and}\\
\textsf{P.B. Medvedev\footnote{Email: medvedev@heron.itep.ru}}\\
\emph{Institute of Theoretical and Experimental Physics,}\\
\emph{B.Cheremushkinskaya st. 25, Moscow, 117218}
}

\date {~}
\maketitle
\thispagestyle{empty}

\begin{abstract}
It has been conjectured that at the stationary point of the
tachyon potential for the non-BPS $\mathrm{D}$-brane or
brane-anti-$\mathrm{D}$-brane pair, the
negative energy density cancels the brane tension. We study this
conjecture using a cubic superstring field theory with
insertion of a double-step inverse picture changing operator. We
compute the tachyon potential at levels $(1/2,\,1)$
and $(2,\,6)$. In the first case
we  obtain that the value of the potential
at the minimum is $97.5\%$ of the non BPS D-brane tension.
Using a special gauge in the second case we get $105.8\%$
of the tension.
\end{abstract}
\newpage
\pagenumbering{arabic}

\tableofcontents

%%%%%%%%%%%%%%%%%%%%%%%%%%%%%%%%%%%%%%%%%%%%%%%%%%%%%%%%%%%%%%%%%
\section{Introduction}
\setcounter{equation}{0}
%\input defenitions.tex
%%%%%%%%%% \section{Introduction}
%\begin{document}

One of the main motivation to construct
string field theory (SFT) \cite{witten1,w2} was a hope to study non-perturbative
phenomena in string theory. SFT gives an off-shell
formulation of a string theory providing a possibility to
investigate non-perturbative phenomena in a systematic way.

The bosonic open string has a tachyon that leads to an instability
of a perturbative vacuum. In the early works by Kostelecky and
Samuel \cite{KS} it was proposed to use SFT to describe
condensation of the tachyon to a stable vacuum. They have shown
that a truncation of open SFT at low levels gives a rather
systematic approximation scheme to calculate the tachyon effective
potential. Moreover, within this calculation scheme the tachyon
potential in open bosonic string has a nontrivial minimum.

Further \cite{AMZ3}, the level truncation method has been applied
to examine an effective potential of auxiliary fields in a cubic
superstring field theory (SSFT) \cite{AMZ1,PTY,AMZ2}. It was found
that some of the low-lying auxiliary scalar fields acquire
non-zero vacuum expectation values providing a new mechanism for
supersymmetry breaking. The gauge vector field becomes massive
while the physical spinor remains  massless, thus the
supersymmetry is broken in the nonperturbative vacuum.

Recently, Sen has proposed  \cite{sen:conjecture} to interpret the
tachyon condensation  as a decay of an unstable D-brane. In the
framework of this interpretation the vacuum energy of the open
bosonic string in the  Kostelecky and Samuel vacuum has to cancel the
tension of the unstable D-brane (more precisely a difference
between Kostelecky and Samuel vacuum and the unstable perturbative
vacuum should be equal to the tension of the unstable D-brane).
In the cubic SFT this cancellation has been checked \cite{9912249} at the low
levels. Further it was argued that the value of the tachyon
potential at the minimum cancels $99\%$ of bosonic
D-brane's  tension \cite{0002237}.

The tachyon potential has been also evaluated in non-polynomial
\cite{9503099} open NS string field theory \cite{0001084}. Note
that the tachyon comes from GSO$-$ sector and
calculations involve only the NS string. Later on  \cite{0002211},
the calculations have been expanded up to the level $3/2$ and
$85\%$ of the non-BPS D-brane's tension has been cancelled. In the
subsequent papers \cite{0003220}  the calculations involving higher
levels were performed and the value of the tachyon potential at
the minimum was found to cancel $90.5\%$ of the brane tension.

In this note we compute the  tachyon potential at low levels of
cubic SSFT. We also find nontrivial minimum with the
value of potential being $105.8\%$ of the brane tension.

The paper is organized as follows. Section \ref{sec:SSFT} contains a brief
review of cubic SSFT.
In Section \ref{sec:calc} the actual calculations
of the tachyon potential
up to the second nontrivial level are performed.
In Section \ref{sec:tension} we determine
the brane tension in cubic theory.
Appendices
contain necessary information and proof of the odd bracket properties .

%\end{document}

%%%%%%%%%%%%%%%%%%%%%%%%%%%%%%%%%%%%%%%%%%%%%%%%%%%%%%%%%%%%%%%%%
\section{Superstring Field Theory on the Branes}
\label{sec:SSFT}
\setcounter{equation}{0}
%\input definitions.tex
%%%%%%%%%% \section{Introduction}
%\begin{document}

\subsection{Cubic Super String Field Theory}
The original Witten's proposal \cite{w2} for NSR superstring
field theory action reads (we list only NS sector, relevant for
the following):
\begin{equation}
S_{W}\cong \int \mathcal{A}\star Q_B \mathcal{A} + \frac23\int
\mathrm{X}\mathcal{A} \star \mathcal{A}\star \mathcal{A}. \label{w}
\end{equation}
Here $Q_B$ is the BRST charge, $\int$ and $\star$ are Witten's
string integral and star product to be specified below. The
string field $\mathcal{A}$ is a series with each term
being a state in the Fock space $\mathcal{H}$ multiplied by space-time
field. States in $\mathcal{H}$ are created by the modes of the
matter fields $X^{\mu}$ and $\psi^{\mu}$, conformal ghosts $b,\,c$
and superghosts $\beta,\,\gamma$:
\begin{equation}
\mathcal{A}=\sum A_{i\dots}(x)\,\beta_i
...\gamma_j...b_k...c_l...\alpha^{\mu}_n... \psi^{\nu}_m
|0\rangle_{-1}. \label{f}
\end{equation}

The characteristic feature of the action (\ref{w}) is the choice
of the $-1$ picture for the string field $\mathcal{A}$. The vacuum
$|0\rangle _{-1}$  in the NS sector is defined as
\begin{equation}
|0\rangle_{-1}=c(0)e^{-\phi(0)}|0\rangle,
 \label{vacuum}
\end{equation}
where $|0\rangle$ stands for $SL(2,\mathbb{R})$-invariant vacuum
and $\phi$ is the field "bosonizing" the $\beta,\,\gamma$ system:
$\gamma=\eta e^{\phi}$, $\beta=e^{-\phi}\partial\xi$. The
insertion of the picture-changing operator \cite{fms}
\begin{equation}
\mathrm{X}=\frac1{\alpha^{\,\prime}} e^{\phi}\psi\cdot\partial
X+c\partial\xi +\frac14 b\partial\eta
e^{2\phi}+\frac14\partial(b\eta e^{2\phi}) \label{picX}
\end{equation}
in the cubic term is just aimed to absorb the unwanted unit of the
$\phi$ charge as only $\langle 0|e^{-2\phi} |0\rangle \neq 0$.

The action (\ref{w}) suffers from the contact term divergencies
\cite{GS,GK,AM,wendt} which arise when a pair of $\mathrm{X}$-s collides in a point.
This sort of singularities appears already at the tree level. To
overcome this trouble it was proposed to change the picture of NS
string fields from $-1$ to $0$, i.e. to replace $|0\rangle_{-1}$
in (\ref{f}) by $|0\rangle$ \cite{AMZ1,PTY}. States in the $-1$
picture can be obtained from the states in the $0$ picture by the
action of the inverse picture-changing operator $Y$ \cite{fms}
$$
Y=4c\partial\xi e^{-2\phi}
$$
with $\mathrm{X}Y=Y\mathrm{X}=1$. This identity holds outside the ranges
$\ker \mathrm{X}$ and $\ker Y$. Therefore at the $0$ picture
there are states that can not
be obtained by applying the picture changing operators
$\mathrm{X}$ and $Y$ to the states at the $-1$ picture.

 The action for the NS string field in the $0$ picture has the
cubic form with the insertion of a double-step inverse
picture-changing operator $Y_{-2}$ \cite{AMZ1,PTY}\footnote{One
can cast the
 action into the same form as the action for the bosonic string if one
 modifies the NS string integral accounting the "measure" $Y_{-2}$:
 $\int^{\prime}=\int Y_{-2}$.}:
\begin{equation}
\label{AMZPTY} S\cong \int Y_{-2}\mathcal{A}\star Q_B\mathcal{A}
 + \frac23\int Y_{-2}\mathcal{A}\star \mathcal{A}\star \mathcal{A}.
\end{equation}
We discuss $Y_{-2}$ in the next subsection.

In the description of the open NSR superstring the string field
$\mathcal{A}$ is
subjected to be GSO$+$. In the $0$ picture there is a variety
of auxiliary fields as compared with the $-1$ picture. These fields are
zero by means of the free equation of motion: $Q_B
\mathcal{A}=0$, but they play a significant role in the
off-shell calculations. For instance, a low level
off-shell NS string field expands as
\begin{align*}
\mathcal{A}\cong\int dk\; \{& u(k) c_1-\frac{1}{2}A_{\mu}(k)
ic_1\alpha _{-1}^{\mu}-\frac{1}{4}B_{\mu}(k) \gamma
_{\frac{1}{2}}\psi_{-\frac{1}{2}}^{\mu}
\\
&+\frac{1}{2}F_{\mu  \nu }(k)c_1\psi _{-\frac{1}{2}}^{\mu} \psi
_{-\frac{1}{2}}^{\nu} +B(k)c_0 +r(k)c_1\gamma _{\frac{1}{2}}\beta
_{-\frac{3}{2}}+ \dots\}e^{ik\cdot X(0,0)}|0\rangle.
\end{align*}
Here $u$ and $r$ are just the auxiliary fields mentioned above.

The SSFT based on the action \eqref{AMZPTY} is free from the
drawbacks of the Witten's action  \eqref{w}. The absence of
contact singularities can be explained shortly. Really, the tree
level graphs are generated by solving the classical equation of
motion by perturbation theory. For Witten's action \eqref{w} the
equation reads
$$
Q_B\mathcal{A}+\mathrm{X}\mathcal{A}\star \mathcal{A} =0.
$$
The first nontrivial iteration ($4$-point function) involving the
pair of $\mathrm{X} A\star A$ vertices produces the contact term
singularity when two of $\mathrm{X}$-s collide in a point. In contrast, the
action \eqref{AMZPTY} yields the following equation
$$
Y_{-2}(Q_B\mathcal{A}+\mathcal{A}\star \mathcal{A}) =0.
$$
Outside the $\ker Y_{-2}$ the operator $Y_{-2}$ can be dropped out
and therefore the interaction vertex does not contain any
insertion leading to singularity. The complete proofs of this fact
can be found in \cite{AMZ1,PTY,AMZ2}.

\subsection{Double Step Inverse Picture Changing Operator}
To have well defined SSFT \eqref{AMZPTY}
the double step inverse picture changing
operator must be restricted to be

a) in accord with the identity\footnote{We assume that this
equation is true up to BRST exact operators.}:
\begin{equation}
Y_{-2}\mathrm{X} =\mathrm{X}Y_{-2}=Y,\label{condition}
\end{equation}

b) BRST invariant: $[Q_B,Y_{-2}]=0,$

c) scale invariant conformal field, i.e. conformal weight of
$Y_{-2}$ is $0$,

d) Lorentz invariant conformal field, i.e. $Y_{-2}$ does not
depend on momentum.

\noindent The point a) provides the formal equivalence between the
improved \eqref{AMZPTY} and the original Witten \eqref{w}
actions. The points b) and d) are obvious. The point c) is
necessary to make the insertion of $Y_{-2}$ compatible with the
$\star$-product.

As it was shown in paper \cite{PTY}, there are two (up to BRST
equivalence) possible choices for the operator $Y_{-2}$.

The first operator, the chiral one \cite{AMZ1}, is built
from holomorphic fields in the upper half plane and is given by
\begin{equation}
Y_{-2}(z)=-4e^{-2\phi(z)}-\frac{16}{5\alpha^{\,\prime}}
e^{-3\phi}c\partial\xi\psi\cdot\partial X(z). \label{chiral}
\end{equation}
The identity \eqref{condition} for this operator reads
$$
Y_{-2}(z)\mathrm{X}(z)= \mathrm{X}(z)Y_{-2}(z) =Y(z).
$$
This $Y_{-2}$ is uniquely defined by the constraints a) -- d).

The second operator, the nonchiral one \cite{PTY}, is built from
both holomorphic and antiholomorphic fields in the upper half
plane
 and is of the form:
\begin{equation}
Y_{-2}(z,\overline{z})= Y(z) \widetilde{Y}(\overline{z})
\label{nonchiral}.
\end{equation}
Here by $\widetilde{Y}(\overline{z})$ we denote the
antiholomorphic field $4\tilde{c}(\overline{z})
\overline{\partial}\tilde{\xi}(\overline{z})e^{-2\tilde{\phi}(\overline{z})}$,
where $\tilde{c}(\overline{z}),\,\tilde{\xi}(\overline{z})$ and
$\tilde{\phi}(\overline{z})$ are antiholomorphic ghosts of the
NSR superstring.
For this choice of $Y_{-2}$ the identity \eqref{condition} takes
the form:
$$
Y_{-2}(z,\overline{z})\mathrm{X}(z)= \mathrm{X}(z)Y_{-2}(z,\overline{z})
=\widetilde{Y}(\overline{z}).
$$

The issue of equivalence between the theories based on chiral or
nonchiral insertions still remains open. The first touch to the
problem was performed in \cite{UZ}. It was shown that the actions
for low-level space-time fields are different depending on
insertion being chosen.

In the present paper the actual calculations started in Section
\ref{sec:calc}  are based on \textit{nonchiral}
operator $Y_{-2}(z,\overline{z})$, but up to the end of current
section general points of the discussion are insensitive to the
concrete choice of $Y_{-2}$.

\subsection{SSFT in the Conformal Language}
\begin{figure}[t]
\begin{center}
\includegraphics[width=300pt]{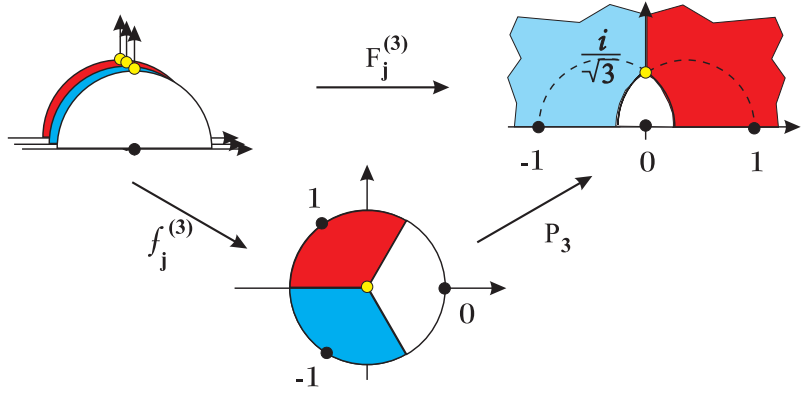}
 \caption{The maps for Witten's vertex.} \label{fig:maps}
\end{center}
\end{figure}

For SSFT calculations it is convenient to employ the tools
of CFT \cite{l'Clare}. States in $\mathcal{H}$ are created by
action of vertex operators (taken in the origin) on the conformal
vacuum $|0\rangle$. In the conformal language $\int$ and
$\star$-product mentioned above are replaced by the odd bracket
$\langle\!\langle...|\dots\rangle\!\rangle$, defined as follows
\begin{equation}
\begin{split}
\langle\!\langle Y_{-2}|A_1,\dots,A_n\rangle\!\rangle&=
\left\langle\, P_n\circ Y_{-2}(0,0)\,F_1^{(n)}\circ A_1(0)\dots
F_n^{(n)}\circ A_n(0)\right\rangle\\
&=\left\langle\, F_j^{(n)}\circ Y_{-2}(i,\overline{i})\,F_1^{(n)}\circ A_1(0)\dots
F_n^{(n)}\circ A_n(0)\right\rangle
, \quad n=2,3.
\end{split}
\label{oddbracket}
\end{equation}
Here r.h.s. contains $SL(2,\mathbb{R})$\,-invariant correlation
function of CFT. $A_j$~$(j=1,\dots,n)$  are vertex operators and
$\{F^{(n)}_j\}$ is a set of maps from the upper half unit disc to
the upper half plane (see Fig.\ref{fig:maps})
\begin{equation}
\begin{split}
F^{(n)}_j=P_n\circ f^{(n)}_j,\quad f^{(n)}_j(w)=
e^{\frac{2\pi i}{n}(2-j)}\left(\frac{1+iw}{1-iw}\right)^{2/n},
\quad j=1,\dots,n,\quad n=2,3;\\
P_2(z)=i\frac{1-z}{1+z},\quad
P_3(z)=\frac{i}{\sqrt{3}}\frac{1-z}{1+z}.
\end{split}
 \label{maps}
\end{equation}
By $F\circ A(0)$ we denote the conformal transform of $A$ by $F$.
For instance, for a primary field $\mathcal{O}_h(z)$ of weight
$h$, one gets
$F\circ\mathcal{O}_h(0)=[F^{\,\prime}(0)]^h\mathcal{O}_h(F(0))$.

$Y_{-2}$ is the double-step inverse picture changing operator
\eqref{nonchiral} inserted in the center of the unit disc. This
choice of the insertion point is very important, since all the
functions $f_j^{(n)}$ maps the points $i$ (the middle points of
the individual strings) to the same point that is the origin.
In other words, the origin is a unique common point for
all strings (see Figure \ref{fig:maps}). The next important
fact is the zero weight of the operator $Y_{-2}$, so its conformal
transformation is very simple $f\circ
Y_{-2}(z,\overline{z})=Y_{-2}(f(z),\overline{f}(\overline{z}))$.
Due to this property it can be inserted in any string.
This note shows that the definition \eqref{oddbracket} is
self-consistent and does not depend on a choice of a string on
which we insert $Y_{-2}$.

Due to the Neumann boundary conditions, there is a relation between
holomorphic and antiholomorphic fields. So it is convenient to
employ a doubling trick (see details in \cite{polchinski}).
Therefore, $Y_{-2}(z,\overline{z})$ can be rewritten in the
following form\footnote{Because of this formula, the operator
$Y_{-2}(z,\overline{z})$ is sometimes called bilocal.}
$$
Y_{-2}(z,\overline{z})=Y(z)Y(z^*).
$$
Here $Y(z)$ is the holomorphic field and $z^*$ denotes the conjugated
point of $z$ with respect to a boundary, i.e. for the unit disc
$z^*=1/\overline{z}$ and $z^*=\overline{z}$ for the upper half
plane. From now on  we work only on the whole complex
plane. Hence the odd bracket takes the form
\begin{equation} \langle\!\langle
Y_{-2}|A_1,\dots,A_n\rangle\!\rangle=\left\langle\,
Y(P_n(0))Y(P_n(\infty))\,F_1^{(n)}\circ A_1(0)\dots F_n^{(n)}\circ
A_n(0)\right\rangle. \label{oddbracket2}
\end{equation}

To summarize, the action we start with reads
\begin{equation}
S[\mathcal{A}]=-\frac{1}{g_{o}^2}\left[\frac12\langle\!\langle
Y_{-2}|\mathcal{A},Q_B\mathcal{A}\rangle\!\rangle +\frac13\langle\!\langle
Y_{-2}|\mathcal{A},\mathcal{A},\mathcal{A}\rangle\!\rangle\right],
\label{action0}
\end{equation}
where $g_o$ is a dimensionless coupling constant.
In Section \ref{sec:tension}
 it will be
related to a tension of a $\mathrm{D}$-brane.

\subsection{Superstring Field Theory on non-BPS $\mathrm{D}$-brane}
To describe the open string states living on a
single non-BPS $\mathrm{D}$-brane one has to add
GSO$-$ states \cite{sen:9904207}.
GSO$-$ states are Grassman even, while
GSO$+$ states are Grassman odd (see Table \ref{tab:1}).
\begin{table}[!h]
\begin{center}
\renewcommand{\arraystretch}{1.4}
\begin{tabular}[h]{||C|c|C|c||}
\hline \textrm{Name}& Parity & \textrm{GSO} & Comment\\ \hline
\hline \mathcal{A}_+ & odd & + & string field in GSO$+$ sector\\
\hline \mathcal{A}_- & even & - & string field in GSO$-$ sector\\
\hline \Lambda_+ & even & + & gauge \\ \cline{1-3} \Lambda_- & odd
& - & parameters\\ \hline
\end{tabular}
\end{center}
\vspace{-0.5cm}\caption{Parity of string fields and gauge
parameters in the 0 picture.}\label{tab:1}
\end{table}

The unique (up to rescaling of the fields)
gauge invariant action unifying GSO$+$
and GSO$-$ sectors is found to be
\begin{equation}
\begin{split}
S[\mathcal{A}_+,\mathcal{A}_-]&=-\frac{1}{g^2_o}\left[
\frac{1}{2}\langle\!\langle Y_{-2}|\mathcal{A}_+,Q_B\mathcal{A}_+
\rangle\!\rangle+\frac{1}{3}\langle\!\langle
Y_{-2}|\mathcal{A}_+,\mathcal{A}_+,\mathcal{A}_+\rangle\!\rangle
\right.\\
&~~~~~~~~~\left.+\frac{1}{2}\langle\!\langle
Y_{-2}|\mathcal{A}_-,Q_B\mathcal{A}_-\rangle\!\rangle
-\langle\!\langle
Y_{-2}|\mathcal{A}_+,\mathcal{A}_-,\mathcal{A}_-\rangle\!\rangle\right].
\end{split}
\label{action}
\end{equation}
Here the factors before the odd brackets are fixed by the
constraint of gauge invariance, that is specified below, and
reality of the string fields $\mathcal{A}_{\pm}$. Variation of
this action with respect to $\mathcal{A}_+$, $\mathcal{A}_-$
yields the following equations of motion\footnote{We assume that
r.h.s. is zero modulo $\ker Y_{-2}$.}
\begin{equation}
\begin{split}
Q_B\mathcal{A}_++\mathcal{A}_+\star \mathcal{A}_+
-\mathcal{A}_-\star \mathcal{A}_-&=0,\\
Q_B\mathcal{A}_-+\mathcal{A}_+\star \mathcal{A}_-
-\mathcal{A}_-\star \mathcal{A}_+&=0.\\
\end{split}
\label{eqmotion}
\end{equation}
To derive these equations we used the
cyclicity property of the odd
bracket (see \eqref{cyclic}). The action \eqref{action} is invariant under the
gauge transformations
\begin{equation}
\begin{split}
\delta \mathcal{A}_+&=Q_B\Lambda_++[\mathcal{A}_+,\Lambda_+]
+\{\mathcal{A}_-,\Lambda_-\},
\\
\delta\mathcal{A}_-&=Q_B\Lambda_-+[\mathcal{A}_-,\Lambda_+]
+\{\mathcal{A}_+,\Lambda_-\},
\end{split}
\label{gauge}
\end{equation}
where $[\,,]$ ($\{\,,\}$) denotes $\star$-commutator
(-anticommutator). To prove the gauge invariance, it is sufficient to
check the covariance of the equations of motion \eqref{eqmotion}
under the gauge
transformations \eqref{gauge}. A simple calculation leads to
\begin{equation*}
\begin{split}
\delta(Q_B\mathcal{A}_+&+\mathcal{A}_+\star \mathcal{A}_+
-\mathcal{A}_-\star \mathcal{A}_-)\\
&=[Q_B\mathcal{A}_++\mathcal{A}_+\star \mathcal{A}_+
-\mathcal{A}_-\star \mathcal{A}_-,\,\Lambda_+]-
[Q_B\mathcal{A}_-+\mathcal{A}_+\star \mathcal{A}_-
-\mathcal{A}_-\star \mathcal{A}_+,\,\Lambda_-],\\
\delta(Q_B\mathcal{A}_-&+\mathcal{A}_+\star \mathcal{A}_-
-\mathcal{A}_-\star \mathcal{A}_+)\\
&=[Q_B\mathcal{A}_-+\mathcal{A}_+\star \mathcal{A}_-
-\mathcal{A}_-\star \mathcal{A}_+,\,\Lambda_+]+
[Q_B\mathcal{A}_++\mathcal{A}_+\star \mathcal{A}_+
-\mathcal{A}_-\star \mathcal{A}_-,\,\Lambda_-].
\end{split}
\end{equation*}
Note that to obtain this result the associativity of
$\star$-product and Leibnitz rule for $Q_B$ must be employed. These
properties follow from the cyclicity property of the odd bracket
(see Appendix \ref{app:cyclic}).
The formulae above show that the gauge transformations define a Lie algebra.

%\end{document}

%%%%%%%%%%%%%%%%%%%%%%%%%%%%%%%%%%%%%%%%%%%%%%%%%%%%%%%%%%%%%%%%%
\section{Computation of the Tachyon Potential}\label{sec:calc}
\setcounter{equation}{0}
%\documentclass{article}
%\input defenitions.tex

%\begin{document}

Here we explore the
tachyon condensation on the non-BPS $\mathrm{D}$-brane. In the
first subsection, we describe the expansion of the
string field relevant to the tachyon condensation and the level
expansion of the action. In the second subsection we calculate
the tachyon potential up to levels 1 and 4, and find its minimum.

\subsection{The Tachyon String Field}\label{sec:t-field}
The useful devices for computation of the tachyon potential were elaborated in
\cite{9912249,9911116,0001084}. We employ these devices without additional references.

Denote by $\mathcal{H}_1$ the subset of vertex operators of ghost
number $1$ and picture $0$, created by the matter stress tensor
$T_B$, matter supercurrent $T_F$ and the ghost fields $b$, $c$,
$\pd\xi$, $\eta$ and $\phi$. We restrict the string fields
$\mathcal{A}_+$ and $\mathcal{A}_-$ to be in this subspace
$\mathcal{H}_1$. We also restrict ourselves by Lorentz scalars and
put the momentum in vertex operators equal to zero.

Next we expand $\mathcal{A}_{\pm}$ in a basis of $L_0$ eigenstates,
and write the action (\ref{action}) in terms of space-time component
fields. The string field is now a series with each term being
a vertex operator from $\mathcal{H}_1$ multiplied by
a space-time component field. We define
the level $K$ of string field's component $A_i$ to be $h+1$, where
$h$ is the conformal dimension of the vertex operator multiplied by $A_i$,
i.e. by convention the tachyon is taken to have
level $1/2$.
To compose the action truncated at level $(K,\,L)$
we select all the quadratic
and cubic terms  of total level not more than $L$
for the space-time fields of levels not more than $L$.
Since our action is cubic, number
$L$ may be only in range $2K,\dots,3K$.

To calculate the action up to level $(2,6)$ we have
a collection of vertex operators listed in
Table~\ref{tab:2}.
\begin{table}[!t]
\centering
\renewcommand{\arraystretch}{1.4}
\begin{tabular}{||C|C|C|C|C|C||C||}
\hline
\textrm{Level}&\textrm{Weight}&\textrm{GSO}&\textrm{Twist}&\textrm{Name}&
\mathrm{Picture}\;$-1$&\mathrm{Picture}\;$0$\\ \cline{6-7}
L_0+1&L_0&(-1)^{F}&\Omega&&\multicolumn{2}{c||}{\textrm{Vertex operators}}\\
\hline
\hline
0&-1&+&\mathrm{even}&u&\textrm{---}&c\\
\hline
1/2&-1/2&-&\mathrm{even}&t&ce^{-\phi}&e^{\phi}\eta\\
\hline
1&0&+&\mathrm{odd}&r_i&c\partial c\partial\xi e^{-2\phi}&\partial c,\;\;c\partial\phi\\
\hline
3/2&1/2&-&\mathrm{odd}&s_i&c\partial\phi e^{-\phi}&cT_F ,\quad \partial(\eta e^{\phi})\\
&&&&&&bc\eta e^{\phi},\quad \eta\partial e^{\phi}\\
\hline
2&1&+&\mathrm{even}&v_i&\eta,\;\;T_Fce^{-\phi}&\partial^2c ,\;\; cT_B,\;\; cT_{\xi\eta}\\
&&&&&\partial\xi c\partial^2c e^{-2\phi}&cT_{\phi},\;\; c\partial^2\phi ,\;\; T_F\eta e^{\phi}\\
&&&&&\partial^2\xi c\partial c e^{-2\phi}& bc\pd c,\;\; \pd c\pd\phi\\
&&&&&\partial\xi c\partial c \partial e^{-2\phi}& \\
\hline
\end{tabular}
\caption{Vertex operators in pictures $-1$ and $0$.}
\label{tab:2}
\end{table}
Note that there are extra fields in the $0$ picture as compared
with the picture $-1$ (see Section 2.1).
Surprisingly the level $L_0=-1$ is not empty, it contains the field $u$.
One can check that this field is auxiliary. In the following analysis it plays
a significant role. Only due to this field in the next subsection we get
a nontrivial tachyon potential
(as compared with one given in \cite{0004112}) already at level
$(1/2,\,1)$.

As it is shown in Appendix~\ref{app:twist} the string
field theory action in the restricted
subspace $\mathcal{H}_1$ has $\mathbb{Z}_2$ twist symmetry.
Since the tachyon vertex operator
has even twist we can consider a further truncation  of the string field
 by restricting $\mathcal{A}_{\pm}$ to be twist even. Therefore
the fields $r_i ,s_i$ can be dropped out.
Moreover, we impose one more restriction and require our
fields to have $\phi$-charge (see Appendix~\ref{app:notations}) equal to
$0$ and $1$.
String fields (in GSO$\pm$ sectors) up to level $2$ take the
form\footnote{The string fields are presented
without any gauge fixing conditions.}
\begin{equation}
\begin{split}
\mathcal{A}_+(z)&=u\,c(z)+v_1\,\partial^2c(z)+v_2\,cT_B(z)+v_3\,cT_{\eta\xi}(z)
+v_4\,cT_{\phi}(z)\\
&~~~+v_5\,c\partial^2\phi(z)+v_6\,T_F\eta e^{\phi}(z)
+v_7\,bc\pd c(z)+v_8\,\pd c\pd\phi(z),\\
\mathcal{A}_-(z)&=\frac t4\,e^{\phi(z)}\eta(z).
\end{split}
\label{str-f}
\end{equation}

%%%%%%%%%%%%%%%%%%%%%%%%%%%%%%%%%%%%%%%%%%%%%%%%%%%%%%%%%%%%%%
\subsection{The Tachyon Potential}\label{sec:t-potential}
Here we give expressions for the action and the potential by truncating them
up to level $(2,\,6)$.
Since the field (\ref{str-f}) expands over the levels $0$,
$\frac12$, and $2$
we can truncate the action at levels $(1/2,\,1)$ and $(2,\,6)$ only.
All the calculations have been performed on a specially
written program on Maple. All we need is to give to the program the
string fields (\ref{str-f}) and we get the following
lagrangians\footnote{Here we correct the
lagrangian $\mathcal{L}^{(2,\,6)}$ presented in the second
hep-th version of the paper.}
\begin{align}
\mathcal{L}^{(\frac12,\,1)}&=\frac1{g_o^2{\alpha^{\prime}}^{\frac{p+1}2}}
\left[u^2+\frac{1}{4}t^2+\frac{1}{3\gamma^2}ut^2\right],
\label{level1}
\\\mathcal{L}^{(2,\,6)}&=\frac1{g_o^2{\alpha^{\prime}}^{\frac{p+1}2}}
\left[u^2+\frac{1}{4}t^2+(4v_1-2v_3-8v_4+8v_5+2v_7)u\right.
\nonumber
\\&+4v_1^2+\frac{15}{2}v_2^2+v_3^2+\frac{77}{2}v_4^2+22v_5^2+10v_6^2
+8v_1v_3-32v_1v_4+24v_1v_5+4v_1v_7 \nonumber
\\&-16v_3v_4+4v_3v_5-2v_3v_7+12v_3v_8-52v_4v_5-8v_4v_7-20v_4v_8
+8v_5v_7+8v_5v_8 \nonumber
\\&+(-30v_4+20v_5+30v_2)v_6+4v_7v_8\nonumber
\\&+\left(\frac{1}{3\gamma
^2}u+\frac{9}{8}v_1-\frac{25}{32}v_2-\frac{9}{16}v_3-\frac{59}{32}v_4
+\frac{43}{24}v_5+\frac{2}{3}v_7 \right)t^2 \nonumber
\\&\left.
-\left(\frac{40\gamma }{3}u+45\gamma ^3v_1 -\frac{45\gamma^3}{4}v_2
-\frac{45\gamma ^3}{2}v_3-\frac{295\gamma ^3}{4}v_4
+\frac{215\gamma ^3}{3}v_5+\frac{80\gamma^3}{3}v_7\right)v_6^2\right],
\label{level4}
\end{align}
where $\gamma=\frac4{3\sqrt{3}}$.
To simplify the succeeding analysis we use a special gauge choice
\begin{equation}
3v_2-3v_4+2v_5=0.
\label{gauge}
\end{equation}
This gauge eliminates the terms linear in $v_6$ and drastically
simplifies the calculation of the effective potential for the tachyon field.
We will discuss an issue of validity of this gauge in our next paper \cite{ABKMg}.
The effective tachyon potential is
defined as $\mathcal{V}(t)=-\mathcal{L}(t,u(t),v_i(t))$,
where $u(t)$ and $v_i(t)$ are solution to equations of motion
$\partial_{u}\mathcal{L}=0$ and $\partial_{v_i}\mathcal{L}=0$.
In our gauge the equation $\partial_{v_6}\mathcal{L}=0$
admits a solution $v_6=0$ and therefore the tachyon potential
computed at levels $(2,\,4)$ and $(2,\,6)$ is the same.
The potential at levels $(1/2,\,1)$ and $(2,\,6)$
has the following form:
\begin{equation}
\begin{split}
\mathcal{V}_{\mathrm{eff}}^{(\frac12,\,1)}(t)&=\frac{1}{g_o^2\alpha^{\,\prime\,\frac{p+1}{2}}}
\left[ \frac{81}{1024}t^4-\frac{1}{4}t^2
\right],\\
\mathcal{V}_{\mathrm{eff}}^{(2,\,6)}(t)&=\frac{1}{g_o^2\alpha^{\,\prime\,\frac{p+1}{2}}}
\left[ \frac{5053}{69120}t^4-\frac{1}{4}t^2 \right]
\end{split}
\label{potential-g}
\end{equation}
One sees that the potential has two global minima, which are
reached at points $t_c=\pm1.257$ at level $(1/2,\,1)$ and at points
$t_c=\pm 1.308 $ at level $(2,\,6)$ (see also Figure \ref{Fig:potential}
and Table \ref{tab:3}).

The same decreasing of the minima  has a potential calculated at
the  level $(2,\,6)$ with the
fields obtained by applying picture changing operator
\eqref{picX} to the fields in the
$-1$ picture (see Table~\ref{tab:2}).

%\end{document}

%%%%%%%%%%%%%%%%%%%%%%%%%%%%%%%%%%%%%%%%%%%%%%%%%%%%%%%%%%%%%%%%%
\section{Tension of non-BPS $\mathrm{D}p$-brane and Sen's conjecture}
\label{sec:tension}
\setcounter{equation}{0}
%\input definitions.tex

%\begin{document}
%\subsection{Tension of BPS $\mathrm{D}p$-brane from SFT}

To find a tension of  $\mathrm{D}p$-brane following \cite{0002211}
one considers the
SFT describing a pair of $\mathrm{D}p$-branes and calculates the
string field action on a special string field. This string field
\begin{wrapfigure}[11]{l}{55mm}
\vspace{-6mm}
\begin{center}
\includegraphics[width=130pt]{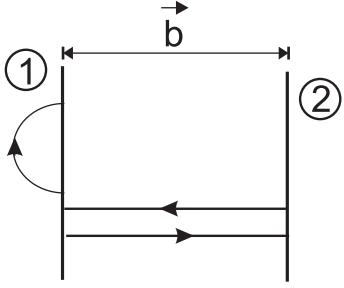}
\end{center}
\vspace{-2mm}
\caption{The system of two non BPS D$p$-branes and
strings attached to them.}
\label{fig:branes}
\end{wrapfigure}
contains a field describing a displacement of one of the branes
and a field describing an arbitrary excitation of the strings
stretched between the two branes. For simplicity one can use
low-energy
excitations of the strings stretched between
the branes.

The cubic SSFT describing a pair of non-BPS $\mathrm{D}p$-branes
includes
 $2\times 2$ Chan-Paton (CP)
factors \cite{0002211,polchinski}
and has the following form
\begin{equation}
\begin{split}
S[\mathcal{\hat{A}}_+,\mathcal{\hat{A}}_-]&=-\frac{1}{g^2_{o}}\left[
\frac{1}{2}\langle\!\langle \hat {Y}_{-2}| \hat {\mathcal{A}}_+,
\hat {Q}_B\hat {\mathcal{A}}_+
\rangle\!\rangle+\frac{1}{3}\langle\!\langle
\hat {Y}_{-2}|\hat {\mathcal{A}}_+,\hat {\mathcal{A}}_+,
\hat {\mathcal{A}}_+\rangle\!\rangle
\right.\\
&~~~~~~~~~\left.+\frac{1}{2}\langle\!\langle
\hat {Y}_{-2}|\hat {\mathcal{A}}_-,\hat {Q}_B\hat {\mathcal{A}}_
-\rangle\!\rangle
-\langle\!\langle
\hat {Y}_{-2}|\hat {\mathcal{A}}_+,\hat {\mathcal{A}}_-,
\hat {\mathcal{A}}_-\rangle\!\rangle\right].
\end{split}
\label{mat-action}
\end{equation}
Here $g_o$ is a dimensionless coupling constant.
The hatted BRST charge $\hat {Q}_B$ and double step inverse picture changing
operator $\hat {Y}_{-2}$ are $Q_B$ and $Y_{-2}$ tensored
by $2\times 2$ unit
matrix.
The string fields are also $2\times2$ matrices
\begin{equation}
\hat{\mathcal{A}}_{\pm}=\mathcal{A}_{\pm}^{(1)}\otimes
\begin{pmatrix}
1 & 0\\
0 & 0
\end{pmatrix}
+
\mathcal{A}_{\pm}^{(2)}\otimes
\begin{pmatrix}
0 & 0\\
0 & 1
\end{pmatrix}
+
\mathcal{B}_{\pm}^*\otimes
\begin{pmatrix}
0 & 1\\
0 & 0
\end{pmatrix}
+
\mathcal{B}_{\pm}\otimes
\begin{pmatrix}
0 & 0\\
1 & 0
\end{pmatrix}
\label{string-fields}
\end{equation}
and the odd bracket includes the trace over matrices.

The action is invariant
under the following gauge transformations:
\begin{equation}
\begin{split}
\delta \hat {\mathcal{A}}_+&=\hat {Q}_B\hat {\Lambda}_+
+[\hat {\mathcal{A}}_+,\hat {\Lambda}_+]
+\{\hat {\mathcal{A}}_-,\hat {\Lambda}_-\},\\ \delta
\hat {\mathcal{A}}_-&=\hat {Q}_B\hat {\Lambda}_-+[\hat {\mathcal{A}}_
-,\Lambda_+]
+\{\hat {\mathcal{A}}_+,\hat {\Lambda}_-\}.
\end{split}
\label{tgauge}
\end{equation}

The fields ${\mathcal A}^{(1)}_{\pm}$
 describe excitations of the string attached to the
first brane, while
$\mathcal{A}^{(2)}_{\pm}$ describe excitations of the string
attached to the second one.
Excitations of the stretched strings are represented by
the fields $\mathcal{B}_{\pm}$ and
$\mathcal{B}_{\pm}^*$ (see Figure~\ref{fig:branes}).
The action for a single non-BPS D-brane (\ref{action}) that we have used above is
derived from the universal action (\ref{mat-action})
by setting
$\mathcal{A}^{(2)}_{\pm}$, $\mathcal{B}_{\pm}$ and $\mathcal{B}^*_{\pm}$
to zero.
Note also that we have not changed the value of the coupling constant $g_o$.

Let us take the following  string fields $\hat{\mathcal{A}}_{\pm}$:
\begin{equation}
\hat{\mathcal{A}}_+=\hat{A}^{(1)}_+ +\hat{B}^*_++\hat{B}_+,
\qquad
\hat{\mathcal{A}}_-=\hat{B}^*_-+\hat{B}_-,
\label{fie}
\end{equation}
where
\begin{align*}
\hat{A}^{(1)}_+&=\int
\frac{d^{p+1}k}{(2\pi)^{p+1}}\;A_i(k_{\alpha})
\mathrm{V}_v^i(k_{\alpha},0)
\otimes
\begin{pmatrix}
1 & 0\\
0 & 0
\end{pmatrix},
\\
\hat{B}_+&=\int\frac{d^{p+1}k}{(2\pi)^{p+1}}\;
iB_{i}(k_{\alpha})
\mathrm{V}_v^{i}(k_{\alpha},
\tfrac{\overrightarrow{b}}{2\pi\alpha^{\prime}})
\otimes
\begin{pmatrix}
0 & 0\\
1 & 0
\end{pmatrix},
\\
\hat{B}^*_+&=\int\frac{d^{p+1}k}{(2\pi)^{p+1}}\;
iB_{i}^* (k_{\alpha})
\mathrm{V}_v^{i}(k_{\alpha},
-\tfrac{\overrightarrow{b}}{2\pi\alpha^{\prime}})
\otimes
\begin{pmatrix}
0 & 1\\
0 & 0
\end{pmatrix}
\\
\hat{B}_-&=\int\frac{d^{p+1}k}{(2\pi)^{p+1}}\;
t(k_{\alpha})
\mathrm{V}_t(k_{\alpha},
\tfrac{\overrightarrow{b}}{2\pi\alpha^{\prime}})
\otimes
\begin{pmatrix}
0 & 0\\
1 & 0
\end{pmatrix},
\\
\hat{B}^*_-&=\int\frac{d^{p+1}k}{(2\pi)^{p+1}}\;
t^*(k_{\alpha})
\mathrm{V}_t(k_{\alpha},
-\tfrac{\overrightarrow{b}}{2\pi\alpha^{\prime}})
\otimes
\begin{pmatrix}
0 & 1\\
0 & 0
\end{pmatrix}
.
\end{align*}
Here $b_i$ is a distance between the branes,
$\alpha=0,\dots,p$ and
$i=p+1,\dots,9$ and
$\mathrm{V}_v^i$ and $\mathrm{V}_t$
are vertex operators of a massless vector and tachyon
fields respectively defined by
\begin{equation}
\begin{split}
\mathrm{V}_v^{\mu}(k_{\alpha},k_i)&=\frac{i}{2}\left[\frac{2}
{\alpha^{\,\prime}}\right]^{1/2}
\left[c\partial X^{\mu}+c2ik\cdot\psi\psi^{\mu}- \frac12\eta
e^{\phi}\psi^{\mu}\right]e^{2ik\cdot X_L(0)},
\\
\mathrm{V}_t(k_{\alpha},k_i)&=\frac{1}{2}
\left[c2ik\cdot\psi- \frac12\eta
e^{\phi}\right]e^{2ik\cdot X_L(0)}.
\end{split}
\label{vec-ver}
\end{equation}
These vertex operators are written in the $0$ picture and
can be obtained
by applying picture changing operator \eqref{picX} to
the corresponding operators in picture $-1$.
The Fourier transform of $A_i(k_{\alpha})$ has an interpretation
of the $\mathrm{D}p$-brane's coordinate up to an overall
normalization factor \cite{polchinski}. Further we will assume that
$b^iB_{i}(k_{\alpha})=0$.

The action for the field \eqref{fie}
depending on the local fields
$B_i(k_{\alpha})$,
$B^*_i(k_{\alpha})$, $t(k_{\alpha})$,
$t^*(k_{\alpha})$ and $A_i(k_{\alpha})$
is given by
\begin{multline}
S[A_i,B_i,t]=-\frac{1}{g^2_o}\left[
\frac{1}{2}\langle\!\langle Y_{-2}|\hat{A}^{(1)}_+,\hat{A}^{(1)}_+\rangle\!\rangle
+\langle\!\langle
Y_{-2}|\hat{B}^*_+,\hat{B}_+\rangle\!\rangle +
\langle\!\langle
Y_{-2}|\hat{B}^*_-,\hat{B}_-\rangle\!\rangle
\right.
\\
\left.+\langle\!\langle Y_{-2}|\hat{A}^{(1)}_+,\hat{B}^*_+,
\hat{B}_+\rangle\!\rangle
-\langle\!\langle Y_{-2}|\hat{A}^{(1)}_+,\hat{B}^*_-,\hat{B}_-
\rangle\!\rangle\right]
=\frac{\alpha^{\,\prime}}{g^2_o
\alpha^{\,\prime\frac{p+1}{2}}}
\int \frac{d^{p+1}k}{(2\pi)^{p+1}}\;
\Biggl\{\frac{k_{\alpha}^2}{2}
A^i(-k)A^i(k)
\\
+B^*_i(-k)B^i(k)\left[k_{\alpha}^2
+\tfrac{b_i^2}{(2\pi\alpha^{\,\prime})^2}\right]
+t^*(-k)t(k)\left[k_{\alpha}^2
+\tfrac{b_i^2}{(2\pi\alpha^{\,\prime})^2}-\tfrac{1}{2\alpha^{\,\prime}}\right]
\\
+\int \frac{d^{p+1}p}{(2\pi)^{p+1}}\;
\gamma^{2\alpha^{\,\prime}(k^2+p^2+p\cdot k+\tfrac{b_i^2}{(2\pi\alpha^{\,\prime})^2})}
\left[B^*_i(-p-k)B^i(k)+\gamma^{-1}t^*(-p-k)t(k)\right]
\frac{b_jA^j(p)}{2\pi\alpha^{\prime}\sqrt{2\alpha^{\prime}}} \Biggr\},
\label{tensc}
\end{multline}
where $\gamma=\frac{4}{3\sqrt{3}}$.
Let us now consider the constant field
$A_i(\xi)=\mathrm{const}$, where $\xi^{\alpha}$ are coordinates
on the brane. Its Fourier transform is of the form
$A_i(p)=(2\pi)^{p+1}A_i\delta(p)$.
Let also $B_i(k)$, $B_i^*(k)$ and
$t(p)$, $t^*(p)$ be on-shell,
i.e. $k_{\alpha}^2+\frac{b_i^2}{(2\pi\alpha^{\,\prime})^2}=0$
and $p_{\alpha}^2+\frac{b_i^2}{(2\pi\alpha^{\,\prime})^2}-\frac{1}{2\alpha^{\,\prime}}=0$.
In this case the action \eqref{tensc} is simplified and takes the form:
\begin{multline}
\tilde{S}[t,B_i]=\frac{\alpha^{\,\prime}}{g_o^2\alpha^{\,\prime\,\frac{p+1}{2}}}
\int_{\text{mass shell}}\frac{d^{p+1}k}{(2\pi)^{p+1}}\;
\left[k_{\alpha}^2
+\frac{b_i^2}{(2\pi\alpha^{\,\prime})^2}-\frac{1}{2\alpha^{\,\prime}}+
\frac{b_i A^i}{2\pi\alpha^{\,\prime}\sqrt{2\alpha^{\,\prime}}}
\right]
t^*(-k)t(k)
\\
+\frac{\alpha^{\,\prime}}{g_o^2\alpha^{\,\prime\,\frac{p+1}{2}}}
\int_{\text{mass shell}}\frac{d^{p+1}k}{(2\pi)^{p+1}}\;
\left[k_{\alpha}^2
+\frac{b_i^2}{(2\pi\alpha^{\,\prime})^2}
+\frac{b_iA^i}{2\pi\alpha^{\,\prime}\sqrt{2\alpha^{\,\prime}}}\right]
B^*_j(-k)B^j(k).
\end{multline}
The field $A_i$ can be interpreted as a shift of the mass.
Therefore one gets the equation
\begin{equation}
\delta(m^2)=\frac{2b_i\delta b^i}{4\pi^2\alpha^{\,\prime\,2}}
=\frac{b_iA^i}{2\pi\alpha^{\,\prime}\sqrt{2\alpha^{\,\prime}}}.
\end{equation}
So one gets
\begin{equation}
A^i=\frac{1}{\pi}\left[\frac{2}{\alpha^{\,\prime}}\right]^{\frac12}
\delta b^i.
\label{chi-b}
\end{equation}
This formula determines the normalization of the field $A_i$.
Therefore we can introduce the profile of the first non-BPS D$p$-brane as
$x_i(\xi)=\pi\left[\frac{\alpha^{\,\prime}}{2}\right]^{1/2}A_i(\xi)$.
Substitution of the $A_i(\xi)$ into the first term of the action
(\ref{tensc}) yields
\begin{equation}
S_0=-\frac{2}{g^2_o\pi^2\alpha^{\,\prime\frac{p+1}{2}}}
\int d^{p+1}\xi\;\frac{1}{2}
\partial_{\alpha}x^i(\xi)\partial^{\alpha}x_i(\xi).
\label{S0X}
\end{equation}
The coefficient before the integral is the non-BPS
D$p$-brane tension $\tilde{\tau}_p$\footnote{
We do not perform here a multiplication of the r.h.s.
of \eqref{tauBPS}
by $\sqrt{2}$ as it was done in the first version of
the paper.}:
\begin{equation}
\tilde{\tau}_p=\frac{2}{g^2_o\pi^2\alpha^{\,\prime\frac{p+1}{2}}}.
\label{tauBPS}
\end{equation}
As compared with the expression for the tension given in \cite{0002211}
we have the addition factor $4$. The origin of this factor is the difference in
the normalization of the superghosts $\beta,\gamma$.

\begin{figure}[t]
\centering
\includegraphics[width=320pt]{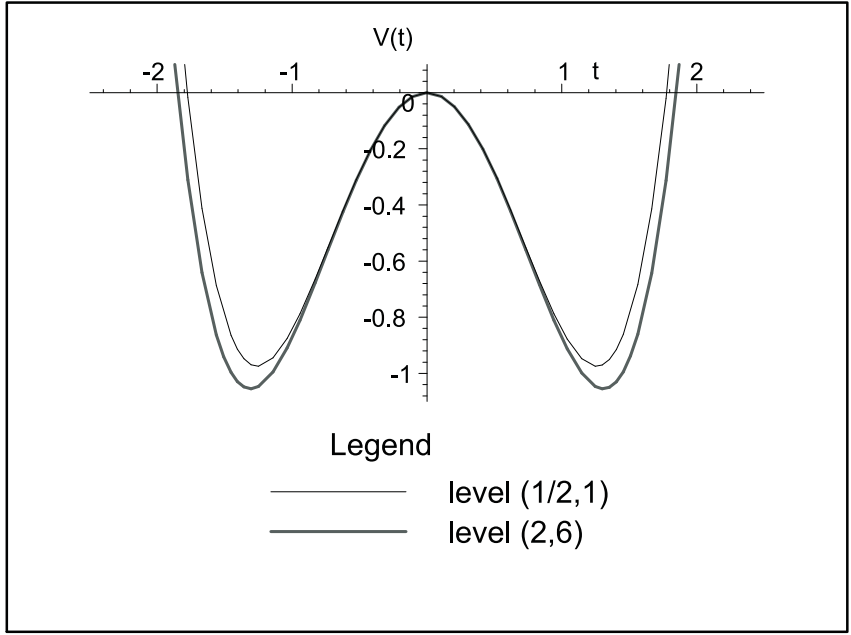}
\caption{Graphics of the tachyon potential at the levels
$(1/2,\,1)$ and $(2,\,6)$.
``$-1$" is equal to the minus tension of non BPS D$p$-brane
$\tilde{\tau}_p=\frac{2}{\pi^2 g^2_o\alpha^{\,\prime\,\frac{p+1}{2}}}$.}
\label{Fig:potential}
\end{figure}

Now we can express the coupling constant $g_o^2$ in terms of the
tension $\tilde{\tau}_p$.
Hence the potential \eqref{potential-g} at levels $(1/2\,1)$ and $(2\,6)$
takes the form (see also Figure \ref{Fig:potential})

\begin{equation}
\begin{split}
\mathcal{V}_{\mathrm{eff}}^{(\frac12,\,1)}(t)&=\frac{\pi^2\tilde{\tau}_p}{2}
\left[ \frac{81}{1024}t^4-\frac{1}{4}t^2
\right],\\
\mathcal{V}_{\mathrm{eff}}^{(2,\,6)}(t)&=\frac{\pi^2\tilde{\tau}_p}{2}
\left[ \frac{5053}{69120}t^4-\frac{1}{4}t^2 \right]
\end{split}
\end{equation}

The critical points of these functions are collected in Table
\ref{tab:3}.
\begin{table}[!thb]
\centering
\renewcommand{\arraystretch}{1.4}
\begin{tabular}{||C|L|L||}
\hline
\textrm{Potential}& \textrm{Critical points} & \textrm{Critical values}\\
\hline
\hline
\mathcal{V}_{\mathrm{eff}}^{(\frac1 2,\,1)} & t_c=0& \mathcal{V}_c=0\\
& t_c=\pm\frac{8\sqrt{2}}{9}\approx \pm1.257&
\mathcal{V}_c\approx-0.975\tilde{\tau}_p\\
\hline
\mathcal{V}_{\mathrm{eff}}^{(2,\,6)} & t_c=0& \mathcal{V}_c=0\\
& t_c=\pm\frac{24}{5053}\sqrt{75795}\approx\pm 1.308&
\mathcal{V}_c\approx-1.058\tilde{\tau}_p\\
\hline
\end{tabular}
\caption{The critical points of the tachyon potential at levels
$(1/2,\,1)$ and $(2,\,6)$.}
\label{tab:3}
\end{table}
One sees that the potential has a global minimum and the value of this
minimum is $97.5\%$ at level $(1/2,\,1)$
and $105.8\%$ at level $(2,\,6)$
of the tension $\tilde{\tau}_p$ of the non-BPS $\mathrm{D}p$-brane.

%\end{document}

%%%%%%%%%%%%%%%%%%%%%%%%%%%%%%%%%%%%%%%%%%%%%%%%%%%%%%%%%%%%%%%%%
\section{Conclusion}
\setcounter{equation}{0}
%\input defenitions.tex
%%%%%%%%%% \section{Introduction}
%\begin{document}

We have computed the effective tachyon potential for
non-BPS-$\mathrm{D}$-brane in cubic SSFT on two first nontrivial
levels. The essential feature of our scheme is the choice of the $0$ picture for
the string field in contrast to the $-1$ picture \cite{0002211}.
This choice of the picture enlarge the set of space-time fields
involved into the calculation at any level.
It is interesting to note that already at the first level the
value of the potential at minimum is
$97.5\%$ of brane's tension.
At the next nontrivial level we use the special gauge \eqref{gauge},
which dramatically simplifies the computations of the tachyon potential.
The validity of this gauge
will be the subject of forthcoming publication \cite{ABKMg}.

In conclusion, our scheme
confirms the existence of the minimum as it was
predicted by Sen's conjecture, and gives
$97.5\%$ of brane's tension at the first step and $105.8\%$ at the second.
Hence, we see that the level truncation scheme does not provide
monotone convergence in this gauge.

%\end{document}

%%%%%%%%%%%%%%%%%%%%%%%%%%%%%%%%%%%%%%%%%%%%%%%%%%%%%%%%%%%
\section*{Acknowledgments}

We would like to thank Oleg Rytchkov for useful discussions
and N.~Berkovits and A.~Sen for
remarks on the first version of this paper.
This work was supported in part
by RFBR grant 99-01-00166 and by RFBR grant for leading scientific
schools. I.A., A.K. and P.M. were supported in part by INTAS grant
99-0590 and D.B. was supported in part by INTAS grant 99-0545.

%%%%%%%%%%%%%%%%%%%%%%%%%%%%%%%%%%%%%%%%%%%%%%%%%%%%%%%%%%%
\newpage
\appendix

\vspace{1cm} \noindent\textbf{\Large Appendix}
\renewcommand{\theequation}{\Alph{section}.\arabic{equation}}

%%%%%%%%%%%%%%%%%%%%%%%%%%%%%%%%%%%%%%%%%%%%%%%%%%%%%%%%%%
\section{Notations}\label{app:notations}
\setcounter{equation}{0}
%\input defenitions.tex
%%%%%%%%%% \section{Introduction}
%\begin{document}

Here we collect
notations we use in our calculations (for more details see
\cite{fms}).
\renewcommand{\arraystretch}{1.5}
\begin{longtable}[h]{||LL||}
\hline
X_L^{\mu}(z)X_L^{\nu}(w)\sim-\frac{\alpha^{\prime}}{2}\eta^{\mu\nu}
\log(z-w)
& \pd X^{\mu}(z)\pd X^{\nu}(w)\sim-\frac{\alpha^{\,\prime}}{2}\eta^{\mu\nu}
\frac{1}{(z-w)^2}
\\
\psi^{\mu}(z)\psi^{\nu}(w)\sim-\frac{\alpha^{\prime}}{2}\eta^{\mu\nu}
\frac{1}{z-w}
&
\\
\hline
\hline
c(z)b(w)\sim b(z)c(w)\sim\frac{1}{z-w}
&
\gamma(z)\beta(w)\sim -\beta(z)\gamma(w)\sim\frac{1}{z-w}
\\
\phi(z)\phi(w)\sim-\log(z-w)
& \xi(z)\eta(w)\sim \eta(z)\xi(w)\sim\frac{1}{z-w}
\\
\langle c(z_1)c(z_2)c(z_3)\rangle=(z_1-z_2)(z_2-z_3)(z_3-z_1)~~~~
&\langle e^{-2\phi(y)}\rangle=1
\\
\multicolumn{2}{||L||}{
\left\langle e^{i2p_{\alpha}X_L^{\alpha}(y)-i2b_{i}X_L^{i}(y)}
e^{i2k_{\beta}X_L^{\beta}(z)+i2b_{j}X_L^{j}(z)}\right\rangle=
(2\pi)^{p+1}\alpha^{\,\prime-\frac{p+1}{2}}\delta(p+k)(y-z)^{2\alpha^{\,\prime}(k^2+b^2)}
}\\
\hline
\hline
T_B=-\frac{1}{\alpha^{\,\prime}}\pd X\cdot\pd X-\frac{1}{\alpha^{\,\prime}}\pd \psi\cdot\psi
& T_F=-\frac{1}{\alpha^{\,\prime}}\pd X\cdot\psi
\\
T_{bc}=-2b\pd c-\pd bc
& T_{\beta\gamma}=-\frac32\beta\pd\gamma-\frac12\pd\beta\gamma
\\
T_{\phi}=-\frac12\pd\phi\pd\phi-\pd^2\phi
& T_{\eta\xi}=\pd\xi\eta
\\
\hline
\hline
\multicolumn{2}{||L||}{
T_B(z)T_B(w)\sim \frac{15}{2(z-w)^4}+\frac{2}{(z-w)^2}T_B(w)+\frac{1}{z-w}\pd T_B(w)}
\\
\multicolumn{2}{||L||}{
T_B(z)T_F(w)\sim\frac{3/2}{(z-w)^2}T_F(w)+\frac{1}{z-w}\pd T_F(w)}
\\
\multicolumn{2}{||L||}{
T_F(z)T_F(w)\sim\frac{5/2}{(z-w)^3}+\frac{1/2}{z-w}T_B(w)}
\\
\hline
\hline
\gamma=\eta e^{\phi}
& \beta=e^{-\phi}\pd\xi
\\
T_{\beta\gamma}=T_{\phi}+T_{\eta\xi}
& \gamma^2=\eta\pd\eta e^{2\phi}
\\
\hline
\hline
\phi-\text{charge:}\quad
\left[\frac{1}{2\pi i}\oint d\zeta\partial\phi(\zeta),
\,\mathcal{A}_q\right]=q\mathcal{A}_q
&
\\
\hline
\hline
\multicolumn{2}{||L||}{Q_B=\frac{1}{2\pi i}\oint d\zeta\,\left[
c(T_B+T_{\beta\gamma}+\frac12 T_{bc})+\frac{1}{\alpha^{\,\prime}}\gamma\psi\cdot\pd X
-\frac14 b\gamma^2\right]}
\\
\multicolumn{2}{||L||}{Q_B=\frac{1}{2\pi i}\oint d\zeta\,\left[
c(T_B+T_{\phi}+T_{\eta\xi}+\frac12 T_{bc})+\frac{1}{\alpha^{\,\prime}}\eta e^{\phi}\psi\cdot\pd X
+\frac14 b\pd\eta\eta e^{2\phi}\right]}
\\
\hline
\hline
\multicolumn{2}{||L||}{
\mathrm{X}=\frac1{\alpha^{\,\prime}} e^{\phi}\psi\cdot\pd X+c\pd\xi
+\frac14 b\pd\eta e^{2\phi}+\frac14\pd(b\eta e^{2\phi})}
\\
Y=4c\pd\xi e^{-2\phi} &\\
\hline
\caption{Notations, correlation functions and OPE-s.}\label{table:2}
\end{longtable}
\renewcommand{\arraystretch}{1}

%\end{document}

%%%%%%%%%%%%%%%%%%%%%%%%%%%%%%%%%%%%%%%%%%%%%%%%%%%%%%%%%%
\section{Cyclicity Property}\label{app:cyclic}
\setcounter{equation}{0}
%\input definitions.tex
%%%%%%%%%% \section{Introduction}
%\begin{document}

The proof of the cyclicity property is very similar to the
one given in \cite{0002211}.
But there is one specific point --- insertion of the
double step inverse picture changing operator $Y_{-2}$
\eqref{nonchiral}.
So we repeat the proof with all necessary modifications.

Let $T_n$ and $R$ denotes rotation
by $-\frac{2\pi}{n}$ and $-2\pi$ respectively:
$$
T_n(w)=e^{-\frac{2\pi i}{n}}w,
\qquad R(w)=e^{-2\pi i}w.
$$
These transformations have two fixed points namely
$0$ and $\infty$. Let us apply the transformation
$T_n$ to the maps $f^{(n)}_k$ \eqref{maps} and
we get the identities:
\begin{equation}
T_n\circ f_{k}^{(n)}=f_{k+1}^{(n)},\quad k<n,\qquad
T_n\circ f_{n}^{(n)}=R\circ f_{1}^{(n)},
\qquad n=2,3.
\end{equation}
Since the weight of the operator $Y_{-2}$
is zero and $0$ and $\infty$ are fixed points of
$T_n$ and $R$,
the operator $Y_{-2}$ remains unchanged.
Due to $SL(2,\mathbb R)$-invariance
of the correlation function
we can write down a chain of equalities
\begin{multline}
\langle Y_{-2}\,F_1^{(n)}\circ A_1\dots
F_{n-1}^{(n)}\circ A_{n-1}F_{n}^{(n)}\circ
\mathcal{O}_h\rangle
=\langle Y_{-2}\,f_1^{(n)}\circ A_1\dots
f_{n-1}^{(n)}\circ A_{n-1}f_{n}^{(n)}\circ
\mathcal{O}_h\rangle
\\
=\langle Y_{-2}\,T_n\circ f_1^{(n)}\circ A_1\dots
T_n\circ f_{n-1}^{(n)}\circ A_{n-1}T_n\circ f_{n}^{(n)}
\circ \mathcal{O}_h\rangle
\\
=e^{-2\pi i h}\langle Y_{-2}\,F_{1}^{(n)}\circ \mathcal{O}_h\,
F_2^{(n)}\circ A_1\dots F_{n}^{(n)}\circ A_{n-1}\rangle.
\end{multline}
In the last line we assume that $\mathcal{O}_h$ is a
primary field of weight $h$
and use the transformation law of primary
fields under rotation:
$$
(R\circ\mathcal{O}_h)(w)=e^{-2\pi ih}\mathcal{O}_h(e^{-2\pi i}w).
$$
Also we change the order of operators in correlation
function
without change of a sign, because
the expression inside the brackets should be odd (otherwise it will be equal to zero)
and therefore no matter whether $\Phi$ odd or even. So the cyclicity property
reads
\begin{equation}
\langle\!\langle Y_{-2}|A_1,\dots,A_n\rangle\!\rangle=e^{-2\pi i h_{n}}\langle\!\langle Y_{-2}|A_n,A_1,\dots,A_{n-1}\rangle\!\rangle
\label{cyclic}
\end{equation}

\medskip
\textsc{Examples.} Now we consider some applications of the cyclicity property (\ref{cyclic}).
GSO$+$ sector consists of the fields with integer weights
and therefore their
exponential factor is equal to $1$,
while GSO$-$ sector consists of the fields with half integer
weights and therefore their
exponential factor is $-1$. Now we give few examples
\begin{subequations}
\label{cyclicGSO-}
\begin{align}
\langle\!\langle Y_{-2}|\mathcal{A}_+,Q_B\mathcal{A}_+\rangle\!\rangle&
=\langle\!\langle Y_{-2}|Q_B\mathcal{A}_+,\mathcal{A}_+\rangle\!\rangle,
\\
\langle\!\langle Y_{-2}|\mathcal{A}_-,Q_B\mathcal{A}_-\rangle\!\rangle&
=-\langle\!\langle Y_{-2}|Q_B\mathcal{A}_-,\mathcal{A}_-\rangle\!\rangle,
\\
\langle\!\langle Y_{-2}|\mathcal{A}_+,\mathcal{A}_-,\mathcal{A}_-\rangle\!\rangle&
=-\langle\!\langle Y_{-2}|\mathcal{A}_-,\mathcal{A}_+,\mathcal{A}_-\rangle\!\rangle
=\langle\!\langle Y_{-2}|\mathcal{A}_-,\mathcal{A}_-,\mathcal{A}_+\rangle\!\rangle.
\end{align}
\end{subequations}

%\end{document}

%%%%%%%%%%%%%%%%%%%%%%%%%%%%%%%%%%%%%%%%%%%%%%%%%%%%%%%%%%
\section{Twist Symmetry}\label{app:twist}
\setcounter{equation}{0}
%\input definitions.tex
%%%%%%%%%% \section{Introduction}
%\begin{document}

The proof of the twist symmetry is similar to the one given in \cite{0002211}.
But there is one specific point --- insertion of the operator $Y_{-2}$.
So we repeat this proof here with all necessary modifications.

A twist symmetry is a relation between correlation functions of operators written in
one order and in the inverse one:
\begin{equation}
\langle\!\langle Y_{-2}|\mathcal{O}_1,\dots,\mathcal{O}_{n}\rangle\!\rangle=(-1)^{?}\langle\!\langle Y_{-2}|\mathcal{O}_n,\dots,\mathcal{O}_{1}\rangle\!\rangle.
\label{twist-problem}
\end{equation}
We are interesting in this relation for $n=3$.

\noindent 1) Let us consider the following transformations
$M(w)=e^{-i\pi}w$ and  $\tilde{I}(w)=e^{i0}/w$.
The transformation $\tilde{I}$ has the following properties:
$$
\tilde{I}(z_1)\tilde{I}(z_2)=\tilde{I}(z_1z_2)\quad
\text{and}\quad
\tilde{I}(z^{2/3})=(\tilde{I}(z))^{2/3}.
$$
The pair of points $0$ and $\infty$ is not affected by
$M$ and $\tilde{I}$, therefore the
double-step inverse picture changing operator $Y_{-2}$ remains unchanged.
For the maps \eqref{maps} we have got the following composition laws
\begin{equation}
f_1^{(3)}\circ M=\tilde{I}\circ f_{1}^{(3)},\quad
f_2^{(3)}\circ M=\tilde{I}\circ f_{3}^{(3)}
\quad\text{and}\quad
f_3^{(3)}\circ M=\tilde{I}\circ f_{2}^{(3)}.
\end{equation}

\noindent 2) Since there is an identity $M\circ\mathcal{O}(0)=e^{-i\pi h}\mathcal{O}(0)$ we can apply
it to (\ref{twist-problem})
\begin{multline}
\langle\!\langle Y_{-2}|\mathcal{O}_1,\mathcal{O}_2,\mathcal{O}_{3}\rangle\!\rangle
=e^{i\pi\sum h_j}\langle Y_{-2}\,f_{1}^{(3)}\circ M\circ\mathcal{O}_1\,
f_{2}^{(3)}\circ M\circ\mathcal{O}_2\,
f_{3}^{(3)}\circ M\circ\mathcal{O}_{3}\rangle
\\
=e^{i\pi\sum h_j}
\langle Y_{-2}\,\tilde{I}\circ f_{3}^{(3)}\circ\mathcal{O}_1
\,\tilde{I}\circ f_{2}^{(3)}\circ\mathcal{O}_2
\,\tilde{I}\circ f_{1}^{(3)}\circ\mathcal{O}_{3}\rangle
\\
=e^{i\pi\sum h_j}
\langle Y_{-2}\,f_{3}^{(3)}\circ\mathcal{O}_1
\,f_{2}^{(3)}\circ\mathcal{O}_2
\,f_{1}^{(3)}\circ\mathcal{O}_{3}\rangle
\end{multline}
in the last line we use the invariance with respect to $SL(2,\mathbb R)$.
Let $N_{odd}$ and $N_{even}$
be a number of \textit{odd} or \textit{even}
respectively fields in the set $\{\mathcal{O}_1,\mathcal{O}_2,\mathcal{O}_3\}$.
After rearranging the fields one gets
\begin{equation}
\langle\!\langle Y_{-2}|\mathcal{O}_1,
\mathcal{O}_2,\mathcal{O}_{3}\rangle\!\rangle
=e^{i\pi\sum h_j}
(-1)^{\frac{N_{odd}\left(N_{odd}-1\right)}{2}}
\langle\!\langle Y_{-2}|\mathcal{O}_3,
\mathcal{O}_2,\mathcal{O}_{1}\rangle\!\rangle.
\label{pretwist1}
\end{equation}

\noindent 3) Since the correlation function
is non zero only for odd expression, number
$N_{odd}$ is odd and
$N_{odd}=2m+1$ for some integer $m$.
Also we have an identity $N_{even}+N_{odd}=3$.
It's not difficult to check that
\begin{equation}
(-1)^{\frac{N_{odd}\left(N_{odd}-1\right)}{2}}=(-1)^m
=(-1)^{\frac{N_{odd}-1}{2}+
\frac{N_{odd}+N_{even}-3}{2}}
=(-1)^{N_{odd}+\frac{N_{even}}{2}}.
\label{pretwist2}
\end{equation}
Combining \eqref{pretwist1} and \eqref{pretwist2} we get the twist property
\begin{equation}
\begin{split}
\langle\!\langle Y_{-2}|\mathcal{O}_1,
\mathcal{O}_2,\mathcal{O}_3\rangle\!\rangle&
=\Omega_1\Omega_2\Omega_3\,
\langle\!\langle Y_{-2}|\mathcal{O}_3,
\mathcal{O}_2,\mathcal{O}_1\rangle\!\rangle,
\\
\mbox{where}\quad\Omega_j&=
\left\{
\begin{tabular}{LL}
(-1)^{h_j+1}, & h_j\in\mathbb Z\quad \mbox{ (i.e. GSO$+$)}\\
(-1)^{h_j+\frac{1}{2}}, & h_j\in \mathbb Z  +\frac12
\quad \mbox{ (i.e. GSO$-$)}.
\end{tabular}
\right.
\end{split}
\label{twist-gen}
\end{equation}

\medskip
\noindent\textsc{Examples.}

\noindent $i)$ Let $A_+=A_1+A_2$. Each term in
$$
A_+^3=A_1^3+(A_1A_2^2+A_2^2A_1)+A_2A_1A_2+(A_2^2A_1+A_1A_2^2)+A_1A_2A_1+A_2^3
$$
should be twist invariant to be nonzero. Therefore we get
$\Omega_1=1$ and $\Omega_2=1$.

\noindent $ii)$ Let we have fields $A_+$ and $A_-=a_1+a_2$.
Using cyclicity property (\ref{cyclicGSO-})
one gets
$$
2(A_+,a_1+a_2,a_1+a_2)=(A_+a_1^2+a_1^2A_+)+(A_+a_2^2+a_2^2A_+)+
(A_+a_1a_2+a_2a_1A_+)+(A_+a_2a_1+a_1a_2A_+).
$$
So we get $\Omega_+=1$ and $\Omega_+\Omega_1\Omega_2=1$. If $\Omega_1=1$ (tachyon's sector)
then $\Omega_2=1$ too. Therefore we
can consider a sector with $\Omega=1$ only.

%\end{document}

%%%%%%%%%%%%%%%%%%%%%%%%%%%%%%%%%%%%%%%%%%%%%%%%%%%%%%%%%%

\end{document}